\documentclass[oldversion]{aa}
\usepackage{epsfig}
\usepackage{graphicx}
\usepackage{txfonts}
\usepackage{natbib}
\usepackage{url}

\bibpunct{(}{)}{;}{a}{}{,} 

\def\ecs{erg~cm$^{-2}$s$^{-1}$}
\def\lum{erg~s$^{-1}$}
\def\bone{4U 0614+09}
\def\btwo{2S 0918-549}

\begin{document}

\title{Relativistic outflow from two \\
  thermonuclear shell flashes on neutron stars}

\authorrunning{in 't Zand, Keek \& Cavecchi}

\author{J.J.M.~in~'t~Zand\inst{1}, L. Keek\inst{2} and Y. Cavecchi\inst{3}}


\institute{     SRON Netherlands Institute for Space Research, Sorbonnelaan 2,
                3584 CA Utrecht, the Netherlands; {\tt jeanz@sron.nl}
           \and
                Center for Relativistic Astrophysics, School of Physics,
                Georgia Institute of Technology, Atlanta, U.S.A.
           \and
                Astronomical Institute 'Anton Pannekoek', University of
                Amsterdam, Science Park 904, 1098 XH Amsterdam, The Netherlands
          }

\date{\it }

\abstract{We study the exceptionally short (32-41 ms) precursors of
  two intermediate-duration thermonuclear X-ray bursts observed with
  the {\it Rossi X-ray Timing Explorer} from the neutron stars in
  \bone\ and \btwo. They exhibit photon fluxes that surpass those at
  the Eddington limit later in the burst by factors of 2.6 to 3.1. We
  are able to explain both the short duration and the super-Eddington
  flux by mildly relativistic outflow velocities of $0.1c$ to $0.3c$
  subsequent to the thermonuclear shell flashes on the neutron
  stars. These are the highest velocities ever measured from any
  thermonuclear flash. The precursor rise times are also exceptionally
  short: about 1 ms. This is inconsistent with predictions for nuclear
  flames spreading laterally as deflagrations and suggests detonations
  instead. This is the first time that a detonation is suggested for
  such a shallow ignition column depth ($y_{\rm ign}\approx
  10^{10}$~g~cm$^{-2}$). The detonation would possibly require a
  faster nuclear reaction chain, such as bypassing the
  $\alpha$-capture on $^{12}$C with the much faster
  $^{12}$C($p,\gamma$)$^{13}$N($\alpha,p$)$^{16}$O process previously
  proposed. We confirm the possibility of a detonation, albeit only in
  the radial direction, through the simulation of the nuclear burning
  with a large nuclear network and at the appropriate ignition depth,
  although it remains to be seen whether the Zel'dovich criterion is
  met. A detonation would also provide the fast flame spreading over
  the surface of the neutron star to allow for the short rise
  times. This needs to be supported by future two-dimensional
  calculations of flame spreading at the relevant column depth. As an
  alternative to the detonation scenario, we speculate on the
  possibility that the whole neutron star surface burns almost
  instantly in the auto-ignition regime. This is motivated by the
  presence of 150 ms precursors with 30 ms rise times in some
  superexpansion bursts from 4U 1820-30 at low ignition column depths
  of $\sim10^8$~g~cm$^{-2}$.

}

\maketitle


\section{Introduction}
\label{sec:intro}

Thermonuclear shell flashes on neutron stars (NSs) heat up the
photospheres to typical temperatures of 10$^7$ K, giving rise to Type
I X-ray bursts
\citep{gri76,woo76,mar77,jos77,swa77,lam78,lew93,stroh06}. The primary
fuel for such flashes (hydrogen and helium) is provided by the
companion star in the hosting low-mass X-ray binary (LMXB).
Currently, about 100 bursting NSs are known in our galaxy, providing a
galactic flash rate of a few per hour. Luminosities often reach
$\sim10^{38}$~\lum, making X-ray bursts easily detectable throughout
the galaxy.

In the past two decades, the harvest of X-ray bursts was rich. This is
due to the Proportional Counter Array (PCA) on RXTE \citep[1995-2012;
  e.g.,][]{jah06,gal08}, the Wide Field Cameras on BeppoSAX
\citep[1996-2002; e.g.,][]{jag97,zan04a} and JEM-X on INTEGRAL
\citep[launched in 2003 and still active; e.g.,][]{lun03}, each
yielding at least 2100 burst detections\footnote{These are collected
  in the 'MINBAR' database \citep{gal10a}, see URL {\tt
    http://burst.sci.monash.edu/minbar}}. The instrument with the
largest photon collecting area is the PCA, at about 8000 cm$^2$,
compared to a few hundred cm$^2$ for the other two instruments. This
yields, for a few-keV blackbody spectrum shining at the Eddington
limit at a distance equal to that of the galactic center, typical
photon rates of 10$^4$~s$^{-1}$. The PCA data, therefore, are
particularly well suited to the study of X-ray bursts at millisecond
timescale.

Most of the X-ray burst signal is due to cooling of the burned
layer. The cooling time, or burst duration, scales with the amount of
cooling matter, therefore with the thickness of the layer
\citep[e.g.,][]{zan14}. Most bursts ignite at a column depth of
$y_{\rm ign}\sim10^8$~g~cm$^{-2}$ and have durations of $\sim1$
min. At the other end of the spectrum are the so-called superbursts
with ignition column depths of $y_{\rm ign} \sim
10^{11-12}$~g~cm$^{-2}$ and durations of $\sim1$~d
\citep{cor00,cum01,stro02a,kee08b}. Bursts that last $\sim1$~hr are
called intermediate duration bursts and have intermediate ignition
column depths of $y_{\rm ign}\sim10^{10}$~g~cm$^{-2}$
\citep{zan05,cum06}. These are of special interest here. They are
thought to arise when there is a relatively thick pile of helium on a
relatively cool NS. On a cold NS, the ignition temperature is reached
deeper in the envelope, ergo the thick ignition layer. Such ignition
conditions are readily found \citep{zan07} in ultracompact X-ray
binaries (UCXBs), in which the NS is accompanied by a hydrogen-poor
helium-rich white dwarf in a compact orbit of period less than
$\sim1$~hr \citep[e.g.,][]{nrj86}. The white dwarf is thought to be
the hydrogen-poor core of a star denuded in the past of its
hydrogen-rich atmosphere by the accretion process. Ignition of thick
helium piles on cool NSs will provide the highest nuclear power.

\begin{figure*}[t]
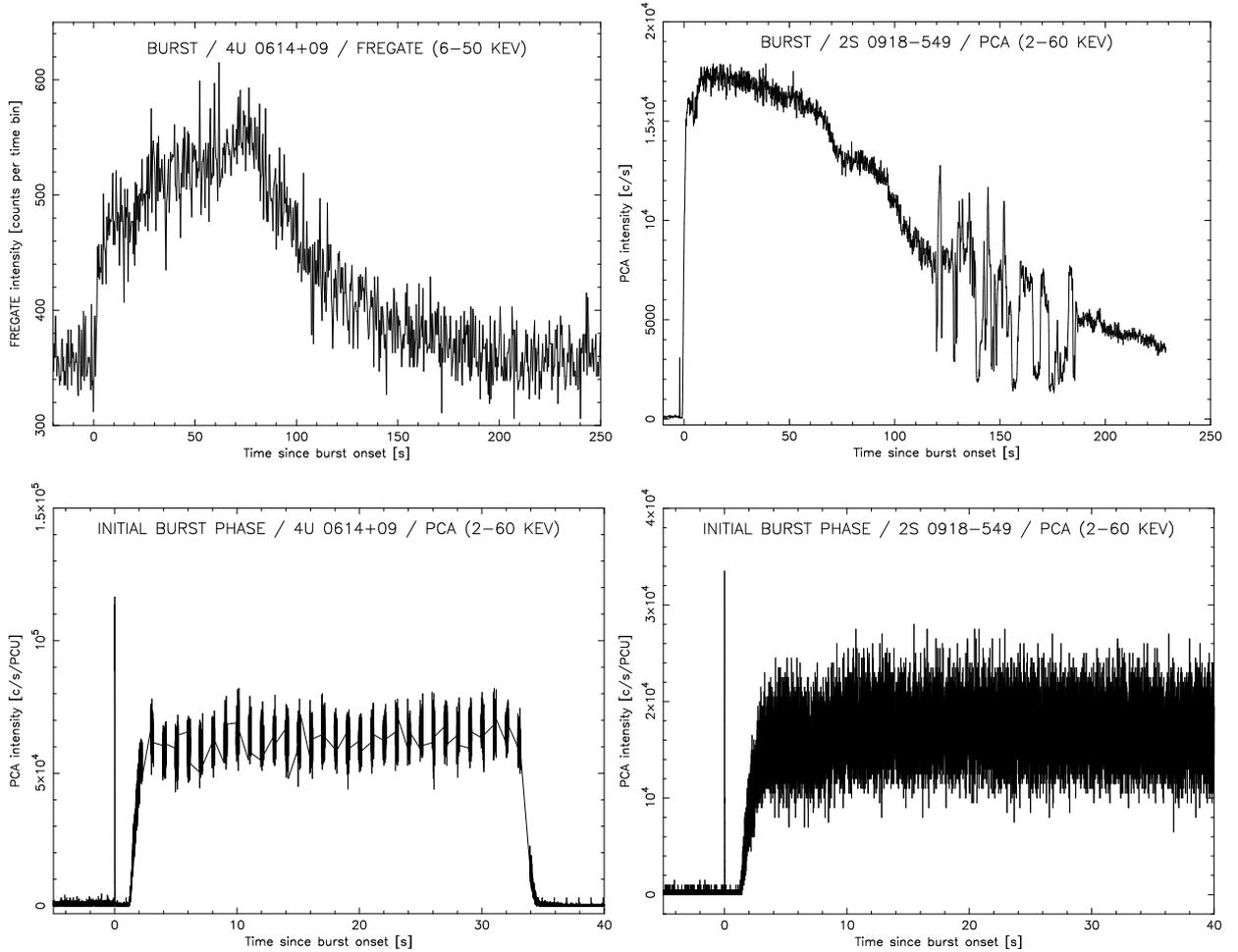

\begin{center}
\includegraphics[height=0.9\columnwidth,angle=270]{precursors_f1a.ps}
\includegraphics[height=0.9\columnwidth,angle=270]{precursors_f1b.ps}
\includegraphics[height=0.9\columnwidth,angle=270]{precursors_f1c.ps}
\includegraphics[height=0.9\columnwidth,angle=270]{precursors_f1d.ps}
\end{center}
\caption{(Top panels) Light curves of the major parts of the bursts,
  left of the burst from \bone\ as measured in 6-50 keV with
  HETE-II-FREGATE \citep{kuu10} and right from \btwo\ as measured in
  2-60 keV with the PCA \citep{zan11}. In both bursts, the end of the
  photospheric expansion phase ('touch-down point') occurs at 75 s,
  recognizable as a peak in the 6-50 keV flux for \bone\ and as a dip
  in the 2-30 keV flux for \btwo. (Bottom panels) Zoomed in light
  curves of the same burst at 2 ms resolution, as measured with the
  PCA. These high-time-resolution measurements suffer from regular
  data gaps due to telemetry saturation. The drop at the end of
  \bone\ is due to the PCA being slewed away from the source. No dead
  time corrections were applied.
  \label{fig:lc0614}}
\end{figure*}

The luminosity during an X-ray burst is determined by the fuel amount
and composition, which dictates specific reaction rates through the
associated nuclear reaction chain. In 20\% of the bursts \citep{gal08}
the luminosity is high enough that it reaches the Eddington
limit. This happens particularly when the energy production is
dominated by helium burning via the 3$\alpha$ process and subsequent
$\alpha$-captures. Although helium burning yields only roughly 1/3 of
the energy per nucleon that hydrogen burning does, the reaction rate
overcompensates this to an extent that the nuclear power from
3$\alpha$ burning is higher than from CNO hydrogen burning
\citep[][]{fuj81,bil98}.  When the power reaches the Eddington limit
\citep[i.e., most nuclear power transforms to radiation, except
  possibly during superbursts, see][]{cum06}, the NS photosphere
expands due to radiation pressure. The photosphere will, at the same
time, cool adiabatically \cite[e.g.,][]{gri80}.  When the expansion is
large enough \citep[we call this 'superexpansion'; ][]{zan10}, the
temperature will move out of the bandpass and the X-ray signal is lost
\citep{taw84b,taw84a,lew84}. The signal returns when the photosphere
moves back to the NS and the superexpansion phase is over. One is left
with the appearance of a precursor. If intermediate duration bursts
are due to helium burning, a high luminosity and superexpansion is
particularly expected since the fuel amount is large and the nuclear
reaction rate high.

There are two hypotheses about the nature of the expanded photosphere
in superexpansion bursts. Both attribute the effect to an expansion to
at least 10$^3$~km, in line with the observation
\citep[e.g.,][]{mol00} that not only the burst radiation fades, but
also the accretion radiation since the X-ray emitting portion of the
accretion disk is covered by the expanded NS photosphere. The first
hypothesis \citep[argued by, e.g.,][]{wal82,ebi83,pac86,jos87,nobi94}
entails a steady-state low-velocity ($\la 0.01c$) wind. \cite{zan10}
noted two problems with this model. First, it does not explain why the
photosphere returns to the NS surface fairly quickly, while the
luminosity does not change as fast and is still Eddington
limited. Second, it does not explain why the superexpansion duration
is independent of burst duration, equivalent to ignition
depth. Therefore, \cite{zan10} propose that something important
happens during the initial stage of the burst, when the quasi-static
wind has not established yet: the expulsion of a shell, like in a
nova, which expands and cools whereby the radiation moves below the
bandpass.  The shell keeps expanding during the 'dark' stage, being
driven by continued Eddington-limited radiation pressure. It dilutes
and, at the end of the superexpansion stage, becomes optically
thin. At this point the NS underneath shines through it, still
radiating at the Eddington limit. The apparent radius is a mix of the
far-away shell, which scatters the NS emission, and the NS and drops
on a timescale of a few seconds. The NS photosphere still radiates at
the Eddington limit and is slightly expanded. This is likely the
quasi-static wind argued by, e.g., \cite{pac86} and \cite{jos87}. It
is less expanded than predicted ($\sim10^{1-2}$ km instead of
$10^{2-3}$~km), possibly because those predictions do not take into
account line driving in a recombined gas expelling the upper cooler
parts of the wind \citep[see][for a more detailed
  discussion]{zan10}. \cite{jos87} show that it takes about 1~s for a
wind to reach a static state. It may be during this dynamic stage that
the geometrically thin optically thick shell is expelled, according to
\cite{zan10}. Interestingly, for a temperature of $\ga1$~GK, the
opacity becomes smaller because it is dominated by Compton instead of
Thompson scattering. As a result, the Eddington limit increases going
into the flash layer by a factor of up to five
\citep{han82}. Therefore, at the start of a burst, a larger fraction
of the layer may be expelled if the nuclear luminosity is larger. The
column thickness of the expelled shell should be at least
$\sim10^4$~g~cm$^{-2}$ to remain optically thick ($>1$~g~cm$^{-2}$) up
to a distance of $\sim10^3$~km. On the other hand, it cannot be
thicker than about 1\% of the ignition column depth, because the
nuclear burning does not provide enough energy \citep[between 1.6 MeV
  and 4.4 MeV per nucleon;][]{fuj87b} to transport more mass out of
the NS gravitational well ($\approx200$ MeV per nucleon for a
canonical NS with mass 1.4~M$_\odot$ and radius 10~km).

Thirty-nine superexpansion bursts have been detected from 9 sources
throughout the 50-year history of X-ray astronomy (see
Appendix~\ref{sebursts}).  More than half are from a single source (4U
1722-30). In almost all cases, precursors last about 1 s. Two
intermediate duration bursts detected with the PCA from \bone\ and
\btwo\ form an exception, with precursors lasting a mere 30-40
ms. These two bursts are the subject of our study. The fast precursors
immediately point to very fast shell velocities and provide
interesting constraints on the physics of ignition, nuclear burning,
flame spreading, and dynamical phenomena of the NS photosphere. In
Sect.~\ref{intro2b} we introduce the general properties of the two
bursts, citing results of previous
studies. Section~\ref{secprecursors} reports the analysis of the
timing and spectral properties of the two precursors. In
Sect.~\ref{discussion} we interpret the properties in terms of the
physical aspects mentioned above. We conclude in
Sect.~\ref{conclusion} and discuss future prospects.

\section{Introducing the two bursts}
\label{intro2b}

\begin{figure*}
\begin{center}
\includegraphics[height=0.9\columnwidth,angle=270]{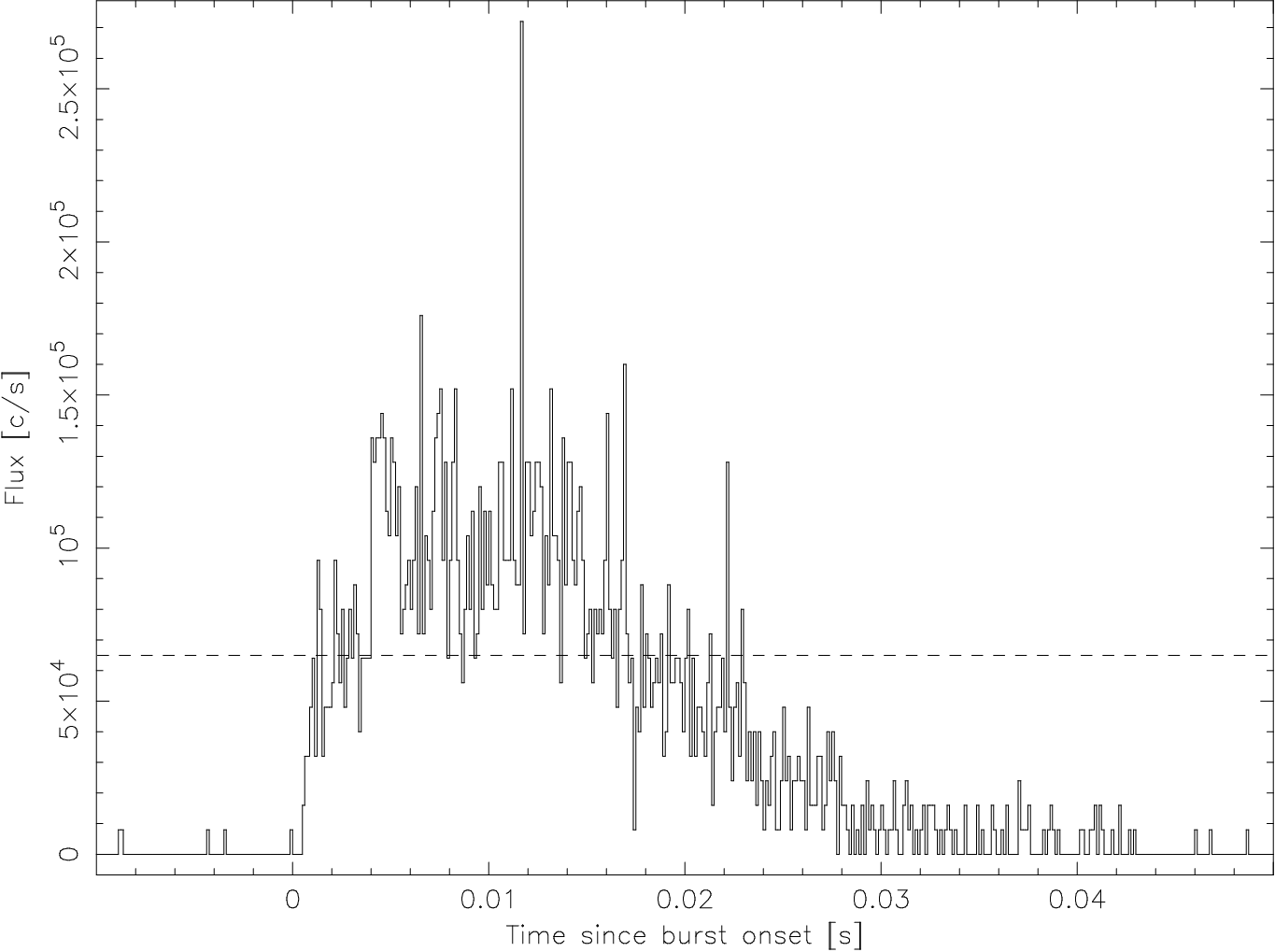}
\includegraphics[height=0.9\columnwidth,angle=270]{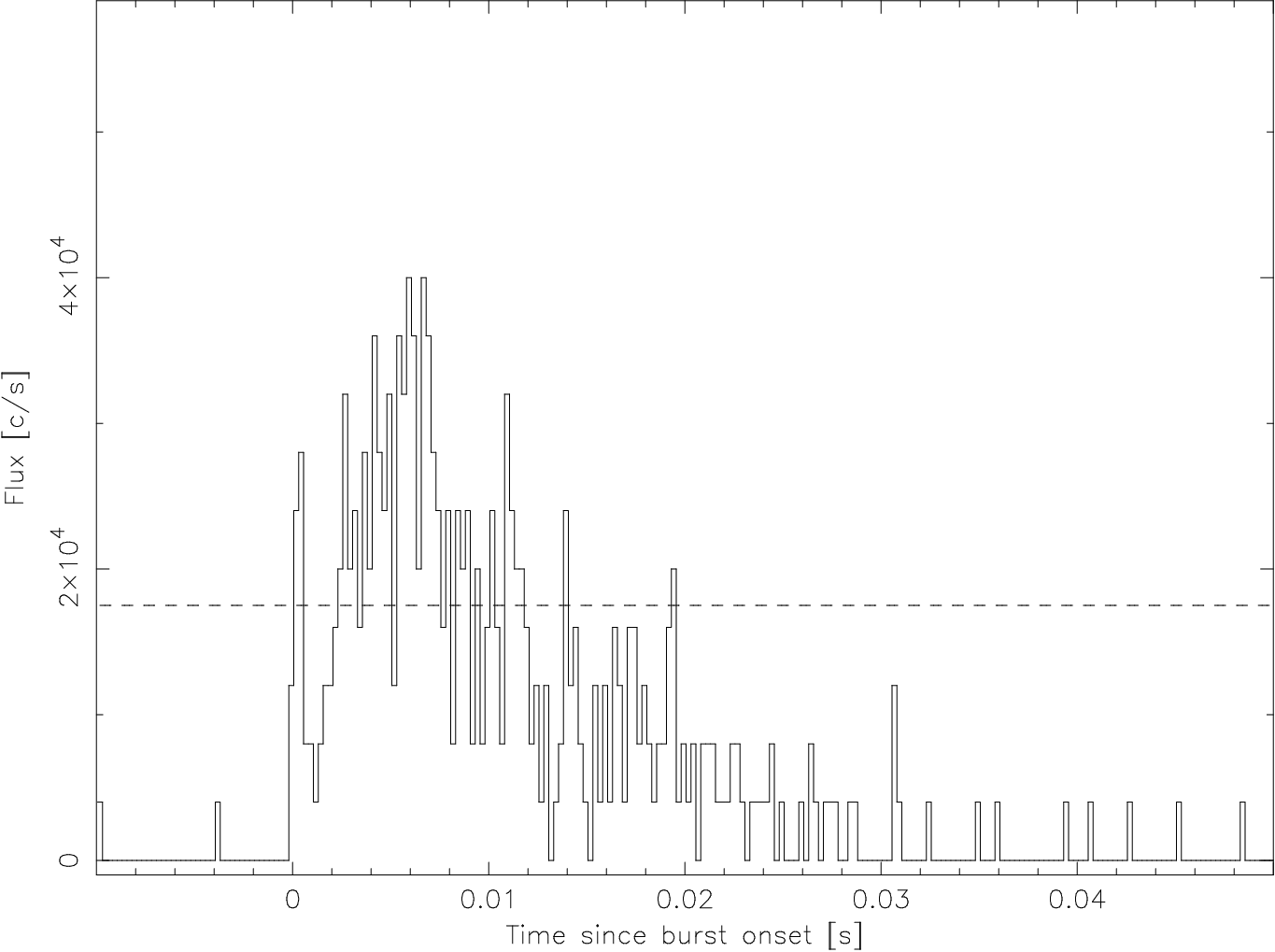}
\includegraphics[height=0.9\columnwidth,angle=270]{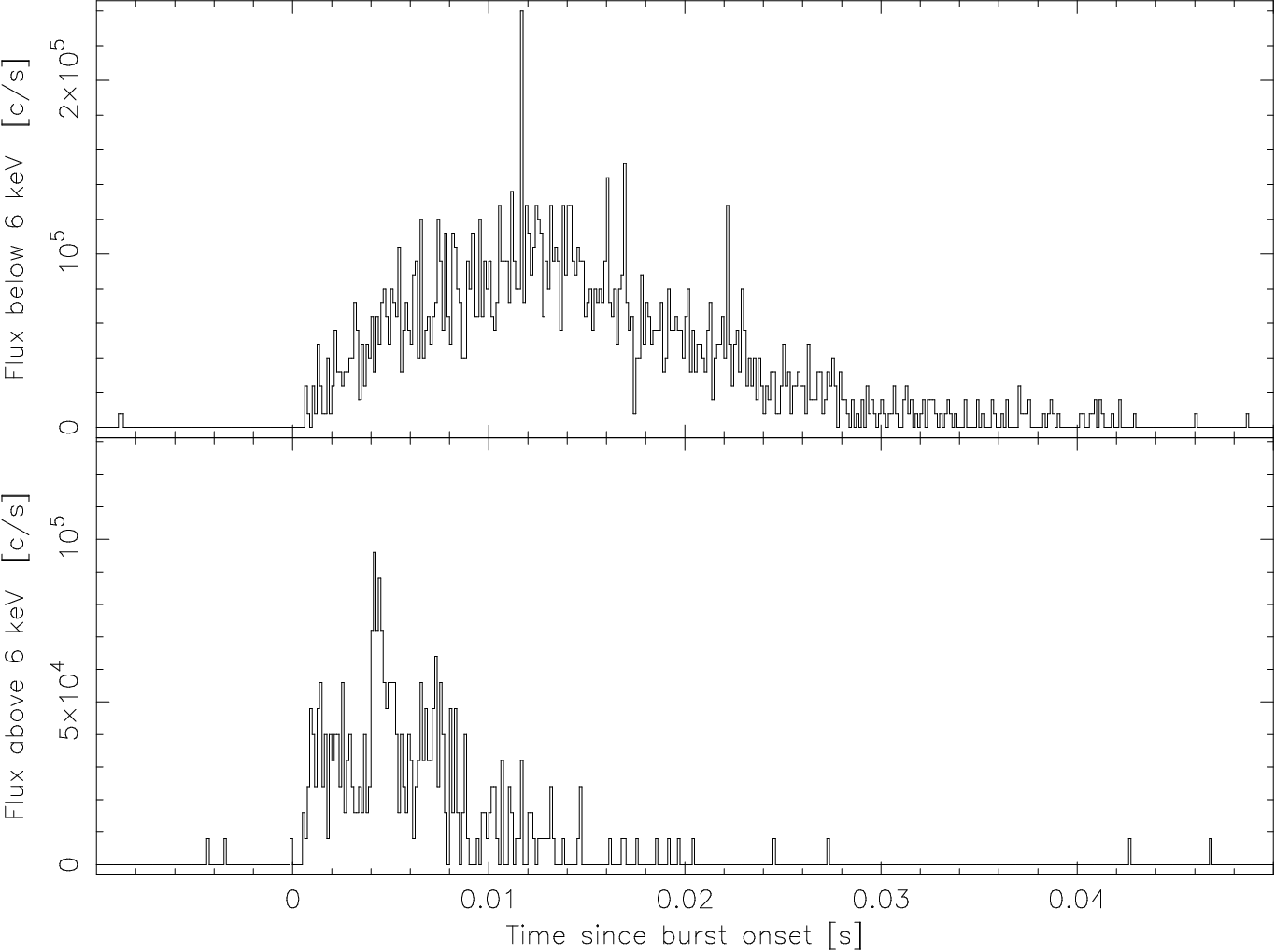}
\includegraphics[height=0.9\columnwidth,angle=270]{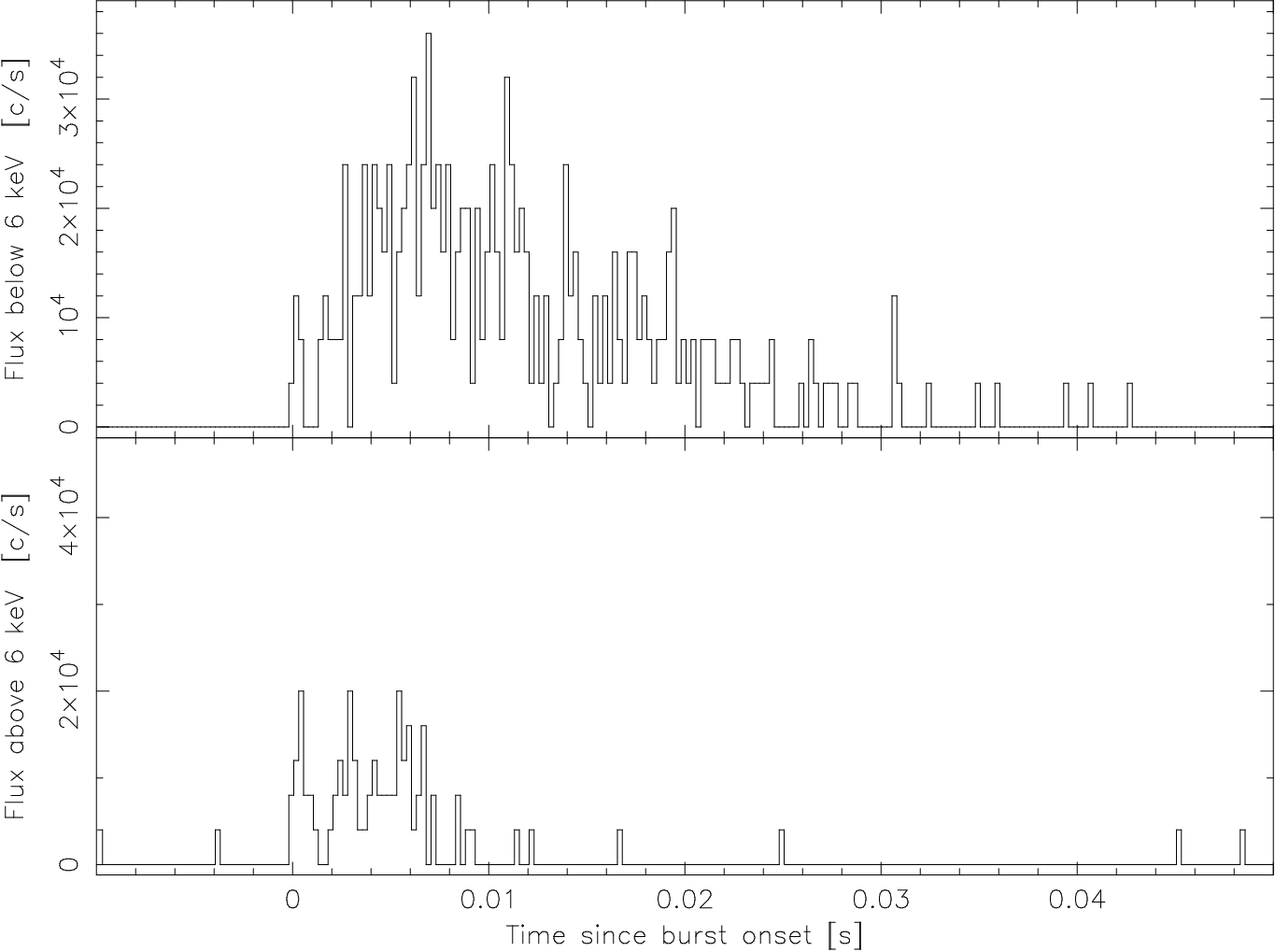}
\end{center}
\caption{X-ray light curves of the precursor to the burst from
  \bone\ at 122 $\mu$s time resolution (left panels) and the burst
  from \btwo\ at 0.25 ms resolution (right panels). The top graphs
  refers to photons at all photon energies, the middle ones to photons
  of energies below 6 keV, and the bottom ones to those above 6
  keV. The X-axis refers to time since burst onset in sec. The
  horizontal dashed lines indicate the level of the flux at which the
  Eddington limit is reached in the main burst.
\label{fig:pre0614}}
\end{figure*}

The initial phases of both bursts were detected with the PCA. The PCA
consists of five co-aligned proportional counter units (PCUs) that
combine to a 8000~cm$^2$ peak effective area at 6 keV in a 2-60 keV
bandpass \citep{jah06}. The spectral resolution is about 20\% (full
width at half maximum) at best and the mostly used data collecting
mode, including here, allows a time resolution of
122~$\mu$s. Generally, not all PCUs operate at the same time.  The
burst light curves are shown in Fig.~\ref{fig:lc0614}. Derived
parameters are listed in Table~\ref{tab1}. Here follows a summary of
the earlier findings.

\begin{table}
\caption[]{Basic parameters of the two bursts (some information from
  \citealt{kuu10} and \citealt{zan11}). \label{tab1}}
\begin{center}
\begin{tabular}{lll}
\hline\hline
Parameter & \bone\ &  \btwo\ \\
\hline
Time (MJD) & 51944.91157 & 54504.12698 \\
Date       & 2001-Feb-04 & 2008-Feb-08 \\
RXTE ObsID & 50031-01-03-05 & 93416-01-05-00 \\
PCUs & 0,1,2,3 & 0,2 \\
Fluence (erg~cm$^{-2}$) & $3.17\times10^{-5}$ & $1.90\times10^{-5}$ \\
Rad. energy outp. (erg) & $3.4\times10^{40}$ & $7\times10^{40}$ \\
Peak flux (\ecs) & $2.66\times10^{-7}$ & $1.18\times10^{-7}$ \\
Timescale (s)$^\dag$ & 120 & 160\\
Precursor duration$^\ddag$ (ms) & 41$\pm$1 & $32\pm1$ \\
Superexpansion duration (s) & 1.15$\pm$0.01 & 1.25$\pm$0.01 \\
Touch-down point (s) & $75\pm5$ & $75\pm5$ \\
Ignition column depth (g~cm$^{-2}$) & 8$\times10^9$ & (1-2)$\times10^{10}$\\
Bol. pers. flux (\ecs) & $3.77\times10^{-9}$ & $5.5\times10^{-10}$ \\
& (1.4\% of peak flux) & (0.5\%)\\
\hline\hline
\end{tabular}
$^\dag$The timescale is defined as the fluence divided by peak flux
\cite[see][]{gal08}; $^\ddag$The precursor duration is measured
between the times that the precursor rises above and decays below 5\%
of the peak flux in the full PCA bandpass.
\end{center}
\end{table}

The source \bone\ contains a NS that has been accreting for at least
40 yr \citep{gia74}. It is an UCXB with an orbital period of probably
50 min \citep{sha08}. There are negligible amounts of hydrogen being
accreted by the NS, as shown by optical spectroscopy \citep{nel06}.
\cite{kuu10} investigated the burst activity since its discovery and
found 30 bursts in almost 40 years of data, with very bright peak
fluxes of up to 15 times the Crab source. Apart from the record holder
Cen X-4 \citep{bel72,kuu09}, this is the brightest of all bursters
which makes it an excellent target for studies at small
timescales. The distance to \bone\ has been determined at 3.2 kpc from
equating the peak flux of Eddington-limited bursts to the Eddington
luminosity limit expected for a hydrogen-poor atmosphere
\citep{bra92,kuu10}. The average accretion rate is low at only 0.8\%
of the H-poor Eddington limit \citep{zan07}. The burst we investigate
here is discussed in \cite{kuu10} and in \cite{zan10}. It is the
brightest burst detected with RXTE. Kuulkers et al. determined an
ignition column depth of $y_{\rm ign}=8\times10^9$~g~cm$^{-2}$ by
modeling the tail of the burst with a cooling envelope.  The burst is
only partly covered by PCA observations, as the observation ended 33 s
after burst onset. The remainder of the burst was measured with other
instruments and, after 1.2 hr, again with the PCA. The duration is
long, with an estimated 6-50 keV e-folding decay time of 40 s. It is a
typical intermediate duration burst \citep[e.g.,][]{zan05,cum06}.

The source \btwo\ is similar to \bone, but at a larger distance
\citep[5.4 kpc;][]{nel04,zan05}. The tentatively measured orbital
period is 17 min \citep{zho10} which would also make it an
ultracompact X-ray binary, as already suspected on the basis of
optical spectroscopy \citep{nel04}. The average accretion rate is
0.5\% of the Eddington limit \citep{zan07}. The burst we discuss was
published previously in \cite{zan11}. The fluence translates to an
energy output of 7$\times10^{40}$~erg for a distance of 5.4 kpc. We
estimate, on the basis of the same method as applied on the burst of
\bone\ \citep[see above and][]{kuu10}, that the ignition column depth
is $y_{\rm ign}=(1-2)\times10^{10}$~g~cm$^{-2}$ (A. Cumming,
priv. comm.). The duration is longer than 310 s, at which time the
observation ended. Therefore, this is a clear intermediate duration
burst from the prototypical source of such bursts \citep{zan05}. Later
on in the burst (120 s to 190 s after burst onset; see
Fig.~\ref{fig:lc0614}) strong upward and downward modulations occur,
which are explained by the effects of an accretion disk which was
dynamically disturbed by the burst outflow and radiation
\citep{zan11}. Another similar example was reported by \cite{deg13}.

The two bursts are quite similar. They are of intermediate duration,
show precursors of similar short duration (32-41 ms), have similar
superexpansion durations (1.15-1.25 s), have similar Eddington-limited
durations (75 s), and arrive from H-poor UCXBs with low accretion rates
onto presumably cool NSs.

\section{The two precursors}
\label{secprecursors}

\subsection{Light curves}

Figure~\ref{fig:pre0614} provides the full details of both precursor
light curves. The data have been collected from four PCUs for
\bone\ and two PCUs for \btwo. Combined with the larger distance for
\btwo, this implies worse statistics for this source.

The following observations can be made from the light curves:

\begin{list}{\leftmargin=0.4cm \itemsep=0cm \parsep=0cm \topsep=0cm}
\item
\item[1.] the precursors last 43 ms (\bone) and 32 ms (\btwo);
\item[2.] both burst rises are just resolved and reach the Eddington
  limit, as determined from the maximum in the main burst phase, very
  quickly - within about 0.5 ms;
\item[3.] the intensities of the bursts surpass the Eddington limit as
  measured in the main burst (see also Fig.~\ref{fig:specs}) by
  factors of 2.6 and 3.1, respectively (after correction for dead
  times in the PCA; see Appendix \ref{sec:deadtime});
\item[4.] the spectrum softens immediately once the Eddington limit is
  first reached, within 0.5 ms from burst onset (as is most easily
  observed by the increasing $<6$~keV intensity and more or less
  constant $>6$~keV intensity);
\item[5.] there is considerable variability on submillisecond
  timescale. For \bone, there is still information contained at the
  maximum time resolution of 122 $\mu$s. It shows a spike 4 ms into
  the burst which lasts 2 ms. Later, after 12 ms, it exhibits an
  unresolved spike which lasts less than 122~$\mu$s. The Poisson
  probability for the flux to rise so high above the local average in
  a single trial is only $6.6\times10^{-6}$. Coincidental or not,
  there appear to be three additional spikes at smaller significance:
  one before and two after the major spike. The most significant one
  has a chance probability for such a high flux or higher of
  $2.7\times10^{-3}$. The three wait times are exactly 5.0 ms. The
  spikes arrive from higher energy photons ($>6$~keV). The data on
  \btwo\ are of less statistical quality, but they also show ms
  variability. There is a marked spike immediately at the start of the
  burst, lasting 1 ms, after which the flux drops by about 75\%.
\end{list}

\subsection{Measurement of expansion speed}
\label{sect:spec}

\begin{figure*}[t]
\includegraphics[width=0.98\columnwidth,angle=0]{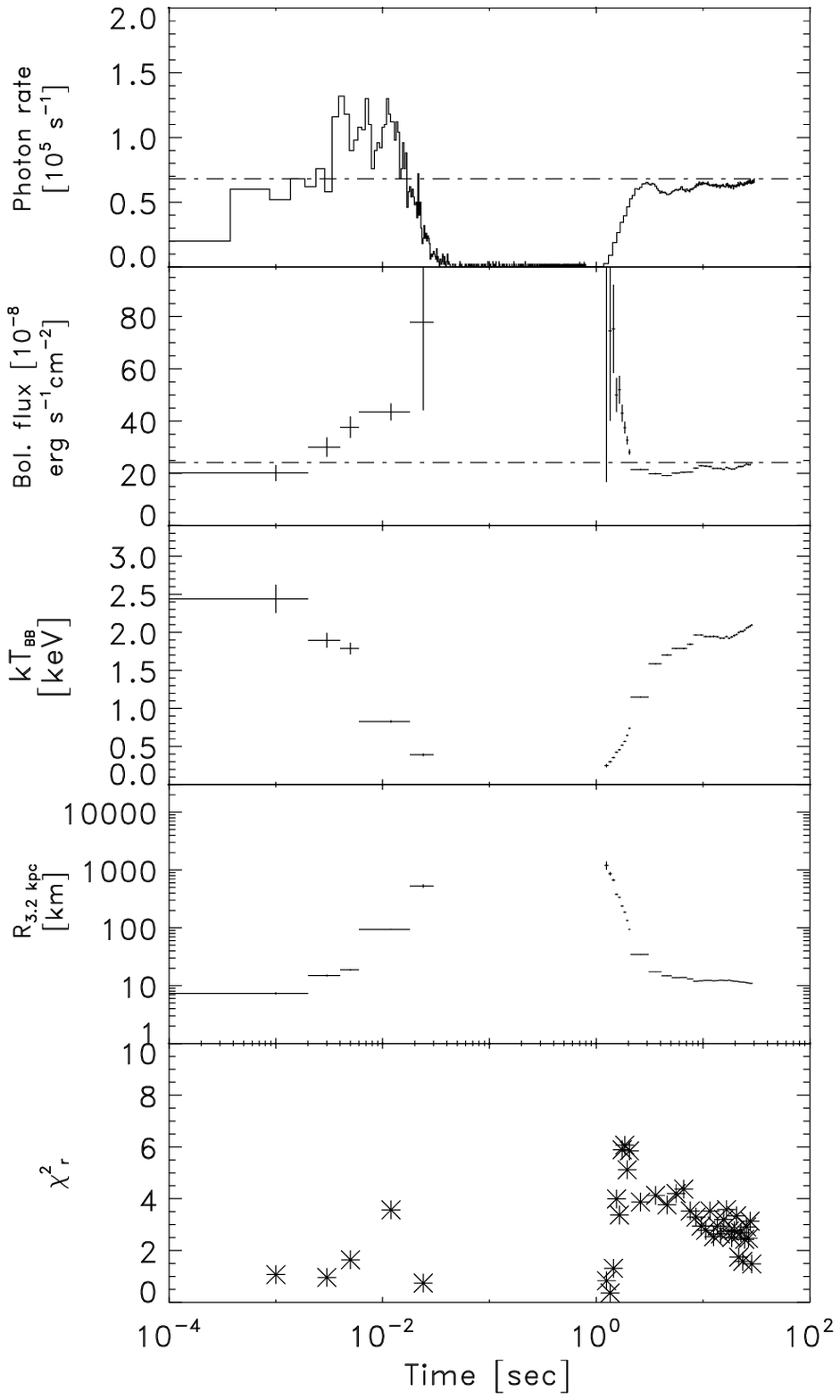}
\includegraphics[width=0.98\columnwidth,angle=0]{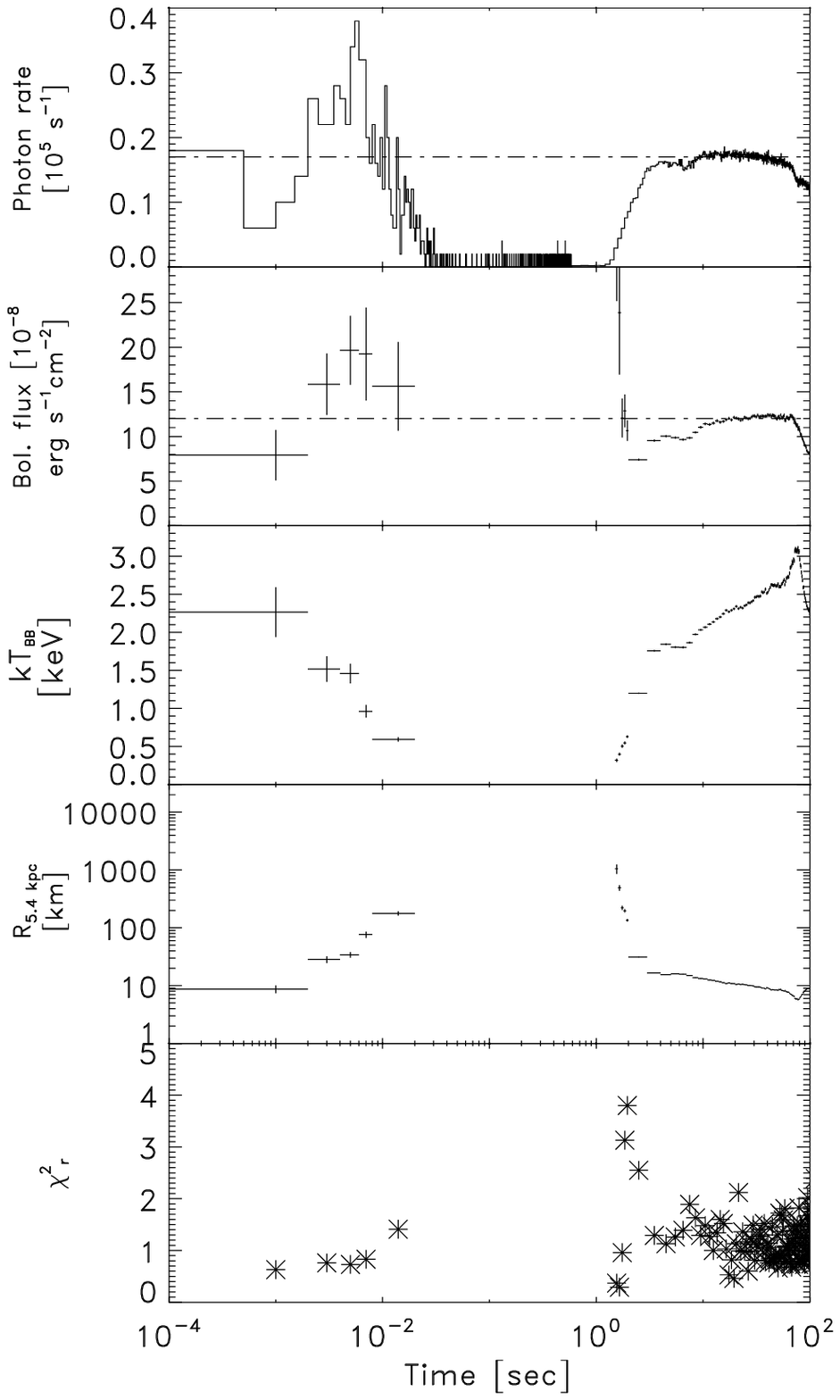}
\caption{Time-resolved spectroscopy of \bone\ (left) and
  \btwo\ (right). The dash-dotted lines indicate maximum flux levels
  measured as late as possible in the burst. The spectral model is a
  simple Planck function. The count rate (top panel) is not corrected
  for dead time while the bolometric flux and radius are. We note that
  the Cash statistic was employed for finding the best fit, not the
  $\chi^2_{\rm r}$ parameter shown in the bottom panels.
\label{fig:specs}}
\end{figure*}

We analyzed the evolving spectrum of the precursors and main bursts.
For each time bin we extracted the well-calibrated \citep{jah06,sha12}
2.8-30 keV spectrum (this involves standard-2 channels 3 and onward,
counting from 0). The spectra were modeled with an absorbed Planck
function, with $N_{\rm H}$ fixed to $3\times10^{21}$ ~cm$^{-2}$ for
both sources \citep{jpc01}, which is actually a negligible amount in
the 2.8-30 keV bandpass. The blackbody temperature and normalization
were fitted using the software package {\tt XSPEC} version 12.8.1g
\citep[e.g.,][]{arn96}. The number of photons is sometimes small and
we decided to employ the Cash statistic \citep{cash} to search for the
best-fit parameter values. We ignored the cosmic and accretion
background since their contribution is negligible (a few tenths of a
percent). We note that the X-ray radiating part of the accretion disk
is blocked out by the expanding photosphere
\citep[e.g.,][]{mol00,zan10}. Figure~\ref{fig:specs} shows the
results. For \bone, the temperature drops to a low value of
$0.391\pm0.018$~keV and the radius increases to $935\pm193$ km before
the signal is lost. This happens within 32 ms and the radius change
translates to an average speed of $(3.9\pm0.8)\times 10
^4$~km~s$^{-1}$ during that time interval. If the evolution of the
radius is studied more closely, an acceleration can be seen. The last
two radius measurements of the precursor translate to a speed of
$(1.1\pm0.3)\times10^5$~km~s$^{-1}$ or $0.3c$.

The apparent speed in \btwo\ is lower than in \bone\ by a factor of
3. The last precursor radius is $242\pm76$~km. With a start value of
10 km, the radius change translates to an average speed of
$(1.7\pm0.5)\times10^4$~km~s$^{-1}$. The radial change between the
last two precursor data points translates to
$(2.9\pm0.9)\times10^4$~km~s$^{-1}$ or $0.1c$.

These speed measurements are uncertain. The large radii are measured
from the Wien tail of the Planck function; the peak of the Planck
function is outside the PCA bandpass. This introduces large
uncertainties in derived temperatures and emission areas. However, the
conclusion that the speed is a few tenths of the speed of light is
justified and supported by the mere fact of the short duration. The
PCA loses the signal from a blackbody if it becomes cooler than 0.25
keV \citep{zan10}. For a constant Eddington luminosity, the equivalent
radius would be larger than 10$^3$~km. If this happens within the
precursor duration of $\approx30$~ms, the average speed must be
$0.1c$. The data clearly point to mildly relativistic outflows.

A caveat that is generally encountered in the analysis of burst
spectra is that they are expected to deviate from the Planck function
because of inverse Compton scattering in the NS atmosphere. This
introduces a systematic difference between the measured 'color'
temperature and the actual effective temperature. Theoretical work
\citep[e.g.,][]{lon86,pav91,sul12} suggests that the effective
temperature is always lower than the color temperature. To arrive at
the same flux, the emission area needs to be larger by approximately
the same factor squared. This implies that our velocities actually
correspond to lower limits.

Although the errors are large, the bolometric flux during the
precursors is seen to increase above the Eddington limit as measured
in the main burst phase after the superexpansion. The bolometric flux
during the precursor of \bone\ is seen to peak at $1.83\pm0.13$ times
the level seen during the main phase. For \btwo\ this is
$1.63\pm0.27$.  These values compare to 3.1 and 2.6, respectively,
found in Sect.~\ref{secprecursors} which apply to the PCA photon count
rate instead of the bolometric flux.

The spectral analysis of the precursor of \bone\ was repeated by
matching the time bins to the most important features of the light
curve. Particularly, we defined time bins that match the flares at 4
and 12 ms. We find that the flare at 4 ms is hotter than the trend
(1.8 versus 1.4 keV at about 3$\sigma$ significance), while the flare
at 12 ms does follow the trend. The temperature accuracies for both
flares are similar.

We note that the spectral analysis is incomplete for the main burst
phase. In contrast to the situation during the precursor phase, the
X-ray radiating part of the accretion disk is thought to be visible
again during the main burst phase and interferes slightly with the
burst spectrum. We do not pursue a full analysis here because we are
interested predominantly in the precursor. For a more complete
analysis of the main burst phase, the reader is referred to
\cite{zan10} and \cite{zan11}. See also \cite{zan13} and \cite{wor13}
for a more extensive study of the behavior of the accretion disk
spectrum during bursts.

\section{Discussion}
\label{discussion}

Our precursor analysis shows three peculiarities: short precursor
durations, super-Eddington fluxes and short rise times. These
observations point to exceptional conditions.  From the following
discussion we conclude that the short rise time measured for the two
bursts is consistent with the helium burning of a thick layer ($y_{\rm
  ign}\ga10^{10}$~g~cm$^{-2}$), in which temperatures rise high enough
to initiate the
$^{12}\mathrm{C}(p,\gamma)^{13}\mathrm{N}(\alpha,p)^{16}\mathrm{O}$
bypass so that probably a detonation is invoked that allows for a fast
spreading of the flame which is unaffected by NS rotation, and a fast
ejection of a shell. Radial velocities are mildly relativistic.

\subsection{Short precursor time and super-Eddington fluxes: relativistic outflow}

The precursors we observe are the shortest for any thermonuclear
flash.  For the 36 other precursor bursts listed in
Appendix~\ref{sebursts} (excluding one superburst), the durations
range between 150 ms and 4 s.  The precursor phenomenon is attributed
to photospheric expansion to radii of 10$^3$~km and beyond
\citep{taw84b,taw84a,lew84}.  The time-resolved spectroscopy of our
bursts (Fig.~\ref{fig:specs}) is consistent with that explanation. Our
precursor durations point to an average speed of $10^3~{\rm
  km}/t_{\rm prec}\approx0.1c$. A close look at the time-resolved
spectroscopy suggests that the speed is not constant and accelerates
to $0.3c$ for \bone. This is a mildly relativistic outflow.

The relativistic character of the outflow allows for a natural
explanation of the second peculiarity, that of super-Eddington fluxes,
provided that the outflow is bulk motion and not, for instance, an
optical depth effect.  The visible approaching side of the shell will
be Doppler boosted by a factor $B$ for the energy flux of
\begin{eqnarray}
B & = & D^{3+\alpha}
\end{eqnarray}
with
\begin{eqnarray}
D=1/\Gamma(1-\beta {\rm cos}\theta),
\end{eqnarray}
$\Gamma$ the Lorentz factor $1/\sqrt{1-\beta^2}$, $\beta=v/c$,
$\theta$ the angle between the velocity and the line of sight and
$\alpha$ the energy spectral index. For a Doppler boost factor equal
to the measured super-Eddington ratio $1.7\pm0.2$ (for the bolometric
luminosity for both sources, see Fig.~\ref{fig:specs}), $\theta=0$,
and $\alpha=0$, $\beta$ would be 0.18$\pm$0.04. This is similar in
magnitude to the estimate from the time-resolved spectroscopy.  It is
worth noting that expansion speeds of a few tenths of the speed of
light are of the same magnitude as the escape velocity from the NS
surface (0.6$c$ for a canonical NS).

Apart from this special relativistic effect, one expects general
relativistic effects as well. The Eddington limit as seen by a distant
observer depends on the photospheric location in the gravitational
well according to
\begin{eqnarray}
F_{\rm Edd,\infty} & \propto & \sqrt{1-\frac{2GM}{Rc^2}}
\end{eqnarray}
with $G$ the gravitational constant, $M$ the NS mass and $R$ the
distance to the NS center of mass \citep[e.g.,][]{dam90}. For a
canonical NS, the Eddington limit would be 31\% smaller for $R=10$~km
than for $R=\infty$. For M$_{\rm NS}=2~$M$_\odot$, this would even be
56\%. These numbers could explain a part of the effect that we
see. However, the same effect should also be visible in other
precursor bursts irrespective of the precursor duration, since it only
depends on $M$ and $R$. We checked this in the literature, see the
references in \cite{zan10} and the light curves in \cite{zan12}, and
found it not to be the case. Furthermore, it should already be clearly
visible in bursts with moderate photospheric expansion. The difference
in the Eddington limit between $R=10$ and $R=20$~km should already be
17\%.  This has not been detected \citep{dam90}. General Relativity
effects do not seem to be a dominant feature in measurements of
$L$. \cite{dam90} attribute this to the dominance of systematic
effects, such as compositional change in the photosphere, changes in
the accretion radiation during the burst and deviations from blackbody
radiation and associated errors in bolometric correction. Furthermore,
the expected changes in $L$ are smaller than what we observe in our
two short precursors.

This would be the first time that Doppler boosting has been detected
in a thermonuclear flash, whether a shell flash on a NS or white
dwarf, or a Type Ia supernova.

\subsection{Short rise time: flash ignition and flame spreading}

The short duration of the precursors goes hand in hand with short rise
times. We find that the rise time to the Eddington limit is $\la 0.5$
ms. The rise time of a burst is expected to be the sum of the nuclear
reaction timescale $t_{\rm nuc}$, the time $t_{\rm wave}$ it takes
the heat wave to travel upward and reach the photosphere, and the time
$t_{\rm spread}$ it takes the flame to spread laterally over the NS
surface. The rise time is determined by the longest of these three
timescales. We discuss each of these three timescales separately.

\subsubsection{$t_{\mathrm{nuc}}$}

\begin{figure}
\includegraphics[width=\columnwidth,angle=0]{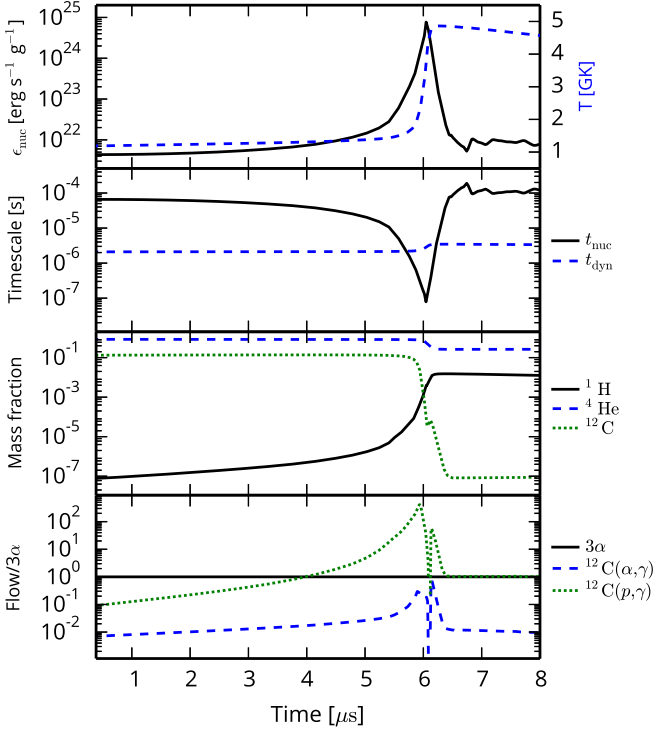}

\caption{\label{fig:burn}Details of the nuclear burning at the start of a
helium flash, around the time when the specific energy generation
rate, $\epsilon_{\mathrm{nuc}}$, is highest. For a KEPLER model in
the zone where runaway thermonuclear burning initiates the flash,
we show as a function of time, $t$, since the start of the local
runaway: $\epsilon_{\mathrm{nuc}}$, temperature $T$ (dashed line),
nuclear burning timescale $t_{\mathrm{nuc}}$, dynamical timescale
$t_{\mathrm{dyn}}$, mass fractions of three isotopes, and the nuclear
flow through several reactions relative to the $3\alpha$ flow. }
\end{figure}

For a helium flash igniting at
$y_{\mathrm{ign}}\simeq10^{10}\,\mathrm{g\, cm^{-2}}$ (pressure
$P\simeq10^{24}\,\mathrm{dyne}\,\mathrm{cm^{-2}}$ for a canonical NS),
burning through the $3\alpha$-process and a chain of $\alpha$-captures
yields $t_{\mathrm{nuc}}\simeq10^{-3}\,\mathrm{s}$
\citep{fuj81}. \citet{wei06}, however, found for
$y_{\mathrm{ign}}=3\times10^{8}\,\mathrm{g\, cm^{-2}}$ that bypassing
the slow $^{12}\mathrm{C}(\alpha,\gamma)^{16}\mathrm{O}$ reaction with
$^{12}\mathrm{C}(p,\gamma)^{13}\mathrm{N}(\alpha,p)^{16}\mathrm{O}$
using protons created in $(\alpha,p)$ reactions substantially reduces
$t_{\mathrm{nuc}}$. One-dimensional simulations performed by
\cite{wal82} with the hydrodynamic stellar evolution code KEPLER
\citep{wea78} of helium flashes with
$y_{\mathrm{ign}}\simeq10^{10}\,\mathrm{g\, cm^{-2}}$ included the
$^{12}\mathrm{C}(p,\gamma)^{13}\mathrm{N}(\alpha,p)^{16}\mathrm{O}$
reactions. Although \cite{wal82} omitted the production of protons by
$(\alpha,p)$ reactions, the importance of this effect became apparent
when hydrogen from their accretion composition mixed into the helium
burning region, and $t_{\mathrm{nuc}}$ was briefly reduced to below 1
$\mu\mathrm{s}$. To further investigate this effect, we performed new
KEPLER simulations of the accretion of $98\%$ by mass
$^{4}\mathrm{He}$ and $2\%$ $^{14}\mathrm{N}$ onto a canonical NS. We
used a modern version of KEPLER that was used in recent X-ray burst
models, and we refer to the respective papers for details
\citep{woo04,heg07a,heg07,kee11}.  Nuclear burning is implemented
using a large adaptive network with thermonuclear rates from a
compilation by \citet{rau03}, which includes the $3\alpha$ reaction,
$\alpha$- and $p$-capture reactions, as well as $(\alpha,p)$
reactions, among others. For our study, we select a burst that ignites
close to the bottom of the accreted fuel column at
$y_{\mathrm{acc}}=1.7\times10^{10}\,\mathrm{g\, cm^{-2}}$, which is on
the order of the values inferred for the bursts discussed in this
paper.

At the burst onset, heating by $3\alpha$ burning creates a convection
region around $y_{\mathrm{ign}}$, and local runaway burning starts
after $16\,\mathrm{s}$ in one zone at $y_{\mathrm{ign}}$. We study the
burning processes in this zone, as we expect $t_{\mathrm{nuc}}$ to be
shortest here (Figure~\ref{fig:burn}). The density in this zone is
$\rho_{\mathrm{ign}}=2.3\times10^{7}\,\mathrm{g\, cm^{-3}}$.  Nuclear
burning before the local runaway increases the temperature to
$T\simeq1\,\mathrm{GK}$, and $(\alpha,p)$ reactions produce a small
number of protons. Even though this number is only $4\times10^{-7}$
times the number of $\alpha$-particles, the
$^{12}\mathrm{C}(p,\gamma)$ reaction already has a much higher rate
than $^{12}\mathrm{C}(\alpha,\gamma)$. The reaction
$^{12}\mathrm{C}(p,\gamma)^{13}\mathrm{N}(\alpha,p)^{16}\mathrm{O}$
bypasses the slower reaction
$^{12}\mathrm{C}(\alpha,\gamma)^{16}\mathrm{O}$, and provides more
seed nuclei for $\alpha-$capture and $(\alpha,p)$ reactions. The
latter reaction further increases the proton mass fraction to as much
as $1.5\%$. At the peak of the specific energy generation rate,
$\epsilon_{\mathrm{nuc}}$, the nuclear flow through
$^{12}\mathrm{C}(p,\gamma)$ exceeds the $3\alpha$ flow by several
orders of magnitude. Most of the $^{12}\mathrm{C}$ is destroyed,
providing seed nuclei for further fast $\alpha$-capture reactions. As
the temperature rapidly increases to $5\,\mathrm{GK}$, there is a
brief dip in the flows of $^{12}\mathrm{C}(p,\gamma)$ and
$^{12}\mathrm{C}(\alpha,\gamma)$, because photo disintegration
enhances the rate of the reverse reactions. Subsequently, nuclear
burning reaches $^{56}\mathrm{Ni}$, which on a longer timescale
undergoes electron capture to form $^{56}\mathrm{Fe}$. This is the
most abundant isotope in the ashes. Finally, the flow through
$^{12}\mathrm{C}(p,\gamma)$ equals the $3\alpha$ flow, as almost all
carbon created by the latter reaction is immediately destroyed by the
former.

We estimate $t_{\mathrm{nuc}}$ from the ratio of the specific internal
energy and $\epsilon_{\mathrm{nuc}}$. At the peak of
$\epsilon_{\mathrm{nuc}}$, it reaches a minimum of
$t_{\mathrm{nuc}}=7.8\times10^{-8}\,\mathrm{s}$. Apart from the
conclusion that it is well below our measured rise times, it has the
important implication that this is shorter than the dynamical
timescale of $t_{\mathrm{dyn}}\simeq3\,\mu s$ (defined as the ratio of
the pressure scale height and the sound speed). In that zone, $t_{\rm
  nuc}$ is shorter than $t_{\rm dyn}$ for $0.49\,\mu\mathrm{s}$ and
within this short time the helium mass fraction is reduced to $0.26$,
while the burning generates $0.92\times10^{18}\,\mathrm{erg\,
  g^{-1}}$.

\subsubsection{$t_{\mathrm{wave}}$}
\label{sec:wave}

Without the
$^{12}\mathrm{C}(p,\gamma)^{13}\mathrm{N}(\alpha,p)^{16}\mathrm{O}$
bypass, burning is subsonic for helium bursts with
$y_{\mathrm{ign}}\simeq10^{10}\,\mathrm{g\, cm^{-2}}$.  Burning
spreads from the ignition depth to lower depths as a deflagration, and
heat is transported towards the surface by convection on a timescale
of $10^{-5}$ s. Previous models with $y_\mathrm{ign}=3\times
10^8\ \mathrm{g\ cm^{-2}}$ \citep[including the bypass
  reaction;][]{wei06} found that convection stalls at a depth where
the thermal timescale is $\sim 1\ \mathrm{ms}$, which set the burst
rise time. For $y_\mathrm{ign}\simeq 10^{10}\ \mathrm{g\ cm^{-2}}$
convection likely reaches even closer to the photosphere, producing a
shorter rise.

Our KEPLER model that includes the
$^{12}\mathrm{C}(p,\gamma)^{13}\mathrm{N}(\alpha,p)^{16}\mathrm{O}$
bypass, however, has hydrodynamic burning, when
$t_{\mathrm{nuc}}<t_{\mathrm{dyn}}$.  For one-dimensional
\citep{wal82} and two-dimensional \citep{zin01} models that satisfy
this condition, the flame spreads as a detonation, and launches a
shock towards the surface. We note that these models require a larger
$y_{\mathrm{ign}}$, as they lack the enhanced energy generation rate
due to the
$^{12}\mathrm{C}(p,\gamma)^{13}\mathrm{N}(\alpha,p)^{16}\mathrm{O}$
bypass. The KEPLER model also includes a shock that reaches the
surface and a shock-breakout peak in the light curve with a timescale
of $\lesssim10^{-6}\,\mathrm{s}$. Fall-back of the shocked outer
atmosphere on a dynamical timescale of
$t_{\mathrm{dyn}}\simeq3\times10^{-6}\,\mathrm{s}$ heats the
atmosphere, which leads to a fast rise of the light curve. Although
the precise details of the shock breakout are likely not accurately
reproduced in our one-dimensional simulation \citep[see,
  e.g.,][]{kee11}, the heating of the photosphere on
$t_{\mathrm{dyn}}$ is robust.

One-dimensional models may not be best suited to determine the
presence of detonation. Detonation requires not only $t_{\rm
  nuc}<t_{\rm dyn}$, but also adherence to the Zel'dovich criterion
\citep[e.g.,][]{zel70} which states that the initial spontaneously
supersonic burning region should be large enough that geometric
dilution does not prematurely terminate the detonation.  \cite{wei07}
apply the Zel'dovich criterion to the case of carbon flashes, but show
that uncertainties remain large and prevent a definite
determination. However, irrespective of whether convective or shock
heating takes place, both produce timescales $t_\mathrm{wave}$ shorter
than the observed rise times of our two bursts.

\subsubsection{$t_{\rm spread}$}

Models proposed for the lateral propagation of the flame fall into two
categories: detonations and deflagrations. In the case of detonations,
after the ignition has started at some location, the flame starts a
shock wave which, advancing, ignites the rest of the fluid via
compression \citep[in different possible ways,
  see][]{zin01,sim12}. The flame proceeds at the same speed as the
shock and can reach supersonic speeds on the order of
$10^9$~cm~s$^{-1}$. Detonations are typically revealed in numerical
simulations of deep helium ignition \citep[$y_{\rm igm} \ga 4\times
  10^{11}$~g~cm$^{-2}$;][]{zin01,sim12}, but so far the rotation of
the star and the effects of the Coriolis force have been
neglected. Such effects may prevent a large bulk motion \citep{cav13}.

On the other hand, deflagrations are driven by thermal conduction
\citep{fryx82,cav13} and proceed more slowly. Even if hydrodynamics,
via the Coriolis force, can increase the speed due to geometrical
effects \citep{spit02,cav13}, velocities remain lower than $\sim
10^6$~cm~s$^{-1}$ \citep{cav13} in the regime of ordinary bursts,
implying that the NS surface is completely covered only after roughly
$t_{\rm spread} = 1$~s.

Our bursts take place at a column depth of $y \sim
10^{10}$~g~cm$^{-2}$, intermediate between the regimes explored so
far.  Our data suggests that the flame spreads over half the NS
circumference in $t_{\rm spread} \la$1 ms. The flame spread time
appears to be out of the realm of deflagrations and in that of
detonations. A tentative conclusion is that we are dealing with
detonations.

However, measurements in a third burster show that this is not so
straightforward.  Short rise times also occasionally appear in bursts
with lower ignition depths. Some superexpansion bursts from 4U 1820-30
have precursors lasting only 0.15 s with rise times of 30 ms (see
Appendix~\ref{sebursts}). These bursts are short, are most likely due
to pure helium ignition \citep{cum03}, have e-folding decay times less
than 10~s and should, therefore, have ignition column depths on the
order of 10$^8$~g~cm$^{-2}$. Therefore, ignition depth is not the only
parameter important for whether the flame propagates as a deflagration
or detonation. Could the NS spin rate be the other one?  The tentative
measurement of a spin of 415 Hz for \bone\ \citep{stroh08} suggests it
is not. The Coriolis force will be strong, but apparently it is not an
issue. This may be explained by the absence of bulk motion, for
instance a flame propagating as a pressure wave in a detonation.

Perhaps we are dealing with neither deflagration nor detonation.  In
the previous section we discuss a simulation showing an extended
convection zone for 16~s at burst onset. The convection is responsible
for the heat transport up to a certain height, creating a much
shallower radial temperature distribution that is near the ignition
condition. The lateral temperature distribution is expected to be
closer to uniform than the radial distribution \cite[e.g.][]{wei07},
so that the fuel layer may be critically close to ignition
throughout. It may not take much lateral heat transport to ignite
neighboring fuel pockets and the ignition may quickly spread over the
NS. This is not a detonation, but it may have similar spread
velocities \citep[this 'auto-' or 'self-ignition' is discussed for
  chemical combustibles in, e.g.,][]{fro93,mak91,bar00}. In this
scenario, the distinguishing factor of bursts with fast rises from
those with slow rises would be the strong lateral homogeneity in the
temperature distribution.

\subsection{The peculiarity of these two bursts}

The question arises why particularly these two bursts have such short
precursors. Apart from this characteristic, the bursts do not seem to
be exceptional. The variety of other bursts with precursors is large,
without a clear trend. They include short and long bursts ($\la1$~min
to 1~hr) and precursors ranging between 0.15 s and 4 s, with no
correlation between them. All bursts with precursors appear to arrive
from hydrogen-deficient UCXBs \citep{zan10,zan12}. Therefore, while
the deficiency of hydrogen seems to be a prerequisite for photospheric
expansion strong enough for precursor appearance, the rapidity of the
precursor must be determined by other parameters. We propose that the
helium abundance is an important such parameter. In 4U 1820-30, the
abundance is high enough \citep{cum03} to generate the needed power
for a fairly fast (0.15 s) expansion ($v=\sim 10^3~{\rm km}/0.15~{\rm
  s}=7\times10^3$~km~s$^{-1}$), even though the ignition column depth
is limited (i.e., on the order of 10$^8$ g~cm$^{-2}$). Perhaps also
the high accretion rate in 4U 1820-30 helps. This makes the start
value of the NS temperature high, so that temperatures during the
flash rise high enough to invoke the fast bypass reaction in the
nuclear reaction chain. Perhaps our two cases are exceptional in
having a high helium abundance as well as a large ignition column
depth. This would provide an exceptionally large luminosity and
radiative driving to push the shell to exceptionally high velocities,
and invoke convection zones large enough to provide short burst
rises. In other long superexpansion bursts \citep[e.g., in 4U 1722-30
  and SLX 1735-269; ][respectively]{mol00,mol05}, the helium abundance
may be smaller.

The duration of the superexpansion stage (i.e., the time between the
onset of the precursor and that of the main burst phase) must be
proportional to the initial column thickness of the shell divided by
the speed. For a duration of 1.2 s and a speed of $0.3c$, the shell
distance to the NS is 10$^5$~km when it becomes optically thin, which
translates to a dilution factor of $10^8$. Therefore, if the speed of
the shell is constant at $0.3c$ ($0.1c$ for \btwo), the initial shell
column thickness would be 10$^8$~g~cm$^{-2}$ (10$^7$~g~cm$^{-2}$)
which is at the limit of the energy constraint of $10^{-2}y_{\rm
  ign}$. Accurate calculations are outside the scope of this paper and
need to take into account the radial structure of the shell (i.e., the
geometric thickness) and the speed evolution, both of which depend on
continued driving by radiation pressure (the flux remains
Eddington-limited for a considerable time after the precursor) and
possibly line driving \citep[e.g.,][]{zan10}.

\subsection{(Sub)millisecond variability}

There is considerable variability in both precursors.  In \bone, we
see after 2.5 ms a spike that lasts 1.5 ms. Notably, this spike is due
to $>6$~keV photons only and the time-resolved spectroscopy shows the
temperature to temporarily increase. In \btwo, we see an initial 1~ms
long spike which subsides within another ms. This is visible in both
bandpasses, in contrast to the spike in \bone, although the
statistical quality precludes a strong statement about
this. Additionally, \bone\ shows flares that are so short that they
cannot be resolved with the 122~$\mu$s resolution of the
data. Interestingly, this is the light-crossing time of a mere 36 km.

The data point to an optically thick outflow.  There is no reason to
believe that this outflow should be isotropic and with one speed. An
irregular structure is plausible, with different pockets of gas
traveling at different speeds. This may result in collisions that give
rise to brief episodes of additional radiation, very much like in the
prompt emission of gamma-ray bursts where internal shocks are thought
to be responsible for the large variability in the prompt emission
\citep{rees94}.

Although our bursts are very bright, no burst oscillations were
detected \citep[for a review of burst oscillations, see][]{wat12}. For
the burst from \bone, this is not a meaningful statement because there
is only PCA data for the Eddington-limited phase. Never has a burst
oscillation been detected during such a phase in any burst. For \btwo,
it is a meaningful statement. A search by \cite{zan11} for burst
oscillations revealed none; the fractional rms upper limit is between
3.9\% and 6.9\% for data stretches of 4 and 1 s, respectively. The
lack of burst oscillations in \btwo\ may be related to the fast flame
spreading preventing the development of a strong enough anisotropy to
give rise to burst oscillations.

\section{Conclusion and future prospect}
\label{conclusion}

We have studied the exceptional onset of two intermediate-duration
thermonuclear X-ray bursts that provide insight into the physics of
flame spreading, nuclear burning and the dynamics of radiatively
driven outflows. We find that the absence of hydrogen and the deep
ignition of helium may yield detonation-like explosions that quickly
traverse the NS radially and laterally, and have large radiative
powers that may result in a relativistic outflow. Better understanding
of this phenomenon needs to come from additional theoretical work and
observations with improved instrumentation.

So far, theoretical work on the outflow focused on the quasi-static
stages later on during the Eddington-limited phase (see
Sect.~\ref{sec:intro}). It would be useful to extend this to the
initial stage that is suspected to be essential for the development of
the shell as an alternative explanation of the superexpansion. In
addition, a theoretical study would be useful of the development of
the structure of the shell, as a means to explain the end stage of
superexpansion (i.e., rise times of the main burst phase and the
variability that is sometimes seen during this rise).  Furthermore, it
would be useful to extend the simulations of ignition to include the
auto-ignition phase, for instance to be able to quantify the
Zel'dovich criterion radially and laterally, and to extend the
simulations of flame spreading at intermediate column depths
($\sim10^{10}$~g~cm$^{-2}$), to investigate in particular the
initiation of detonations and short rise times as a function of
ignition depth, helium abundance, pre-burst temperature, and NS spin
rate.

This study shows that X-ray bursts exhibit significant and interesting
variability on submillisecond timescales that reveal dynamic and
localized phenomena on NS surfaces.  The bursts we study here are
among the top 0.5\% brightest bursts seen thus far, but still only the
largest instrument flown thus far (in the relevant bandpass) enabled
the submillisecond study. With an order-of-magnitude larger photon
collecting area in the same bandpass, more bursts can be studied,
possibly a few tens, some of which can be studied at improved
statistical accuracy. The LOFT mission concept with 10~m$^2$ photon
collecting area \citep{fer12} is excellently suited for these studies.
The envisaged Athena mission with 2~m$^2$ collecting area for the L2
ESA opportunity \citep{nan13} does not have the optimum bandpass (the
effective area at 6~keV is expected to be similar as the PCA on RXTE)
and the observing program will not have as much emphasis on X-ray
bursters, but the soft bandpass will improve the possibility of
measuring the soft spectra at large expansion phases and will better
constrain the photospheric radius than was possible with the
PCA. Furthermore, the higher spectral resolution may reveal narrow
spectral features resulting from absorption by the expelled material
which may be enriched with nuclear ashes. This may particularly be
possible during the initial stages of the main burst when the expelled
material becomes optically thin. Unfortunately, it will be at least
another 12 years before Athena or LOFT may become operational. In the
mean time, ASTROSAT \citep[to be launched within a few
  years;][]{agr06} will provide the best opportunity to continue the
study of superexpansion bursts with a similar combined capability as
RXTE-PCA (through its LAXPC instrument) and Swift-XRT (through its SXT
instrument).

\acknowledgements

We thank Andrew Cumming for calculating the ignition column depth of
the burst from \btwo, Nevin Weinberg and Tullio Bagnoli for useful
discussions, and Craig Markwardt from the RXTE Team for advice on PCA
data analysis. LK acknowledges support from NASA ADAP grant NNX13AI47G
and NSF award AST 1008067.

\bibliographystyle{aa} \bibliography{precursors_refs}

\appendix
\normalsize 
\section{Superexpansion bursts}
\label{sebursts}

Table~\ref{tab2} presents a list of all superexpansion bursts that we
are aware of. This table is an augmented version of Table~1 published
by \cite{zan10}.

\begin{table}[ht]
\caption{List of 39 bursts with superexpansion.\label{tab2}}
\begin{tabular}{l|c|c|c|c|r}
\hline\hline
   & \multicolumn{4}{|c|}{Duration (s)} &  \\
              \cline{2-5}  
Instrument/MJD          &           & Super     & Moderate  &  &Ref.$^\ddag$ \\
             & Precursor & expansion & expansion & \multicolumn{1}{c|}{$\tau_{\rm decay}^\P$} & \\
             & phase     & phase $t_{\rm se}$   & phase $t_{\rm me}$    &  & \\
\hline
\multicolumn{6}{c}{\em 4U 0614+09} \\
\hline
RXTE/51944.903    & 0.04 &  1.2 &  75  &   40(2) &  1 \\
\hline
\multicolumn{6}{c}{\em 2S 0918-549} \\
\hline
RXTE/54504.127    & 0.03 &  1.2 & 75 & 125 & \\ 
\hline
\multicolumn{6}{c}{\em A 1246-588} \\
\hline
WFC/50286.290     &  3.0 &  1.5 &  54   &  38(5)  &  2 \\
WFC/51539.874     &  1.5 &  6.0 &  25   &  19(1)  &  2 \\
\hline
\multicolumn{6}{c}{\em 4U 1708-23 (probably)} \\
\hline
SAS-C/43181.834   &  4.2 &  6.0 &  304  & 300(50) &3,4 \\
\hline
\multicolumn{6}{c}{\em XB 1715-321} \\
\hline
SAS-C/42957.620   &  2.4 & 1.5  &  36   &   30$^\Delta$   &  3 \\
Hakucho/45170.231 &  3   & 4.0  &  105  &   85(5) &  5 \\
\hline
\multicolumn{6}{c}{\em 4U 1722-30} \\
\hline
WFC/50318.279     &  4.0 &  2.0 &   16  &  18(3)  &  6 \\
WFC/50330.196     &  2.0 &  5.0 &   20  &  16(6)  &  6 \\
WFC/50348.938     &  3.0 &  3.0 &   15  &  16(5)  &  6 \\
WFC/50368.307     &  2.0 & 5.0  &   14  &  19(3)  &  6 \\
WFC/50526.311     &  1.0 & 4.0  &   15  &  21(6)  &  6 \\
WFC/50536.895     &  2.5 & 3.5  &   11  &   9(3)  &  6 \\
WFC/50538.439     &  3.0 & 1.0  &   15  &  19(6)  &  6 \\
WFC/50553.130     &  2.5 & 1.5  &   17  &  12(2)  &  6 \\
WFC/50892.706     &  3.0 & 2.5  &   18  &  15(2)  &  6 \\
WFC/50904.813     &  2.5 & 4.0  &   21  &  19(3)  &  6 \\
WFC/51057.579     &  1.5 & 3.0  &   22  &  19(4)  &  6 \\
WFC/51231.379     &  2.0 & 5.0  &   17  &  26(11) &  6 \\
WFC/51270.560     &  2.0 & 5.5  &   25  &  14(2)  &  6 \\
WFC/51278.690     &  2.5 & 2.5  &   15  &  13(1)  &  6 \\
WFC/51422.838     &  2.0 & 6.5  &   22  &  28(5)  &  6 \\
WFC/51431.282     &  4.0 & 5.0  &   27  &  26(5)  &  6 \\
WFC/51453.377     &  1.5 & 5.5  &   20  &  24(6)  &  6 \\
WFC/51461.331     &  4.0 & 3.5  &   19  &  23(4)  &  6 \\
WFC/51610.000     &  3.0 & 3.5  &   20  &  17(3)  &  6 \\
WFC/51639.966     &  1.5 & 5.0  &   32  &  23(5)  &  6 \\
WFC/51956.091     &  2.0 & 3.5  &   41  &  29(10) &  6 \\
RXTE/50395.292    &  3.6 & 1.6  &   23  &  30.2(0.1) &  7 \\
\hline
\multicolumn{6}{c}{\em SLX 1735-269} \\
\hline
I'GRAL/52897.733  &  2.0 & 7.0  &   482 & 600(100)&  8 \\
\hline
\multicolumn{6}{c}{\em 4U 1820-30} \\
\hline
RXTE/51430.074$^\times$    & 15.0 & 2.3  &   1400& 2500    & 9 \\
RXTE/54956.774 & 0.5  & 0.6 & 3.8 & 3.2(1) & 10 \\
RXTE/54958.740 & 0.15 & 1.1 & 3.8 & 3.4(2) & 10 \\
RXTE/54978.321 & 0.15 & 1.3 & 3.8 & 3.7(2) & 10 \\
RXTE/54978.495 & 0.25 & 1.0 & 3.8 & 3.6(2) & 10 \\
RXTE/54981.187 & 0.30 & 0.7 & 3.2 & 3.5(1) & 10 \\
RXTE/54994.534 & 0.60 & 0.7 & 3.2 & 3.1(2) & 10 \\
\hline
\multicolumn{6}{c}{\em M15 X-2} \\
\hline
Ginga/47454.730   &  1.5 &  5.5 &   88  &   60    & 11 \\
WFC/51871.593     &  1.5 & 7.5  &   169 &155(11)  & 12 \\
\hline\hline
\end{tabular}

\noindent 
$^\P$Numbers in parentheses represent 1$\sigma$ uncertainties in the
least significant digit(s); $^\ddag$ 1 - \cite{kuu09}, 2 -
\cite{zan08}, 3 - \cite{hof78}, 4 - \cite{lew84}, 5 - \cite{taw84b}, 6
- \cite{kuu03}, 7 - \cite{mol00}, 8 - \cite{mol05}, 9 -
\cite{stro02a}, 10 - \cite{zan12}, 11 - \cite{jvp90}, 12 -
\cite{zan07} ; $^\Delta$This number is rather uncertain due to
incomplete coverage of the burst; $^\times$superburst

\end{table}

\begin{figure}[t]
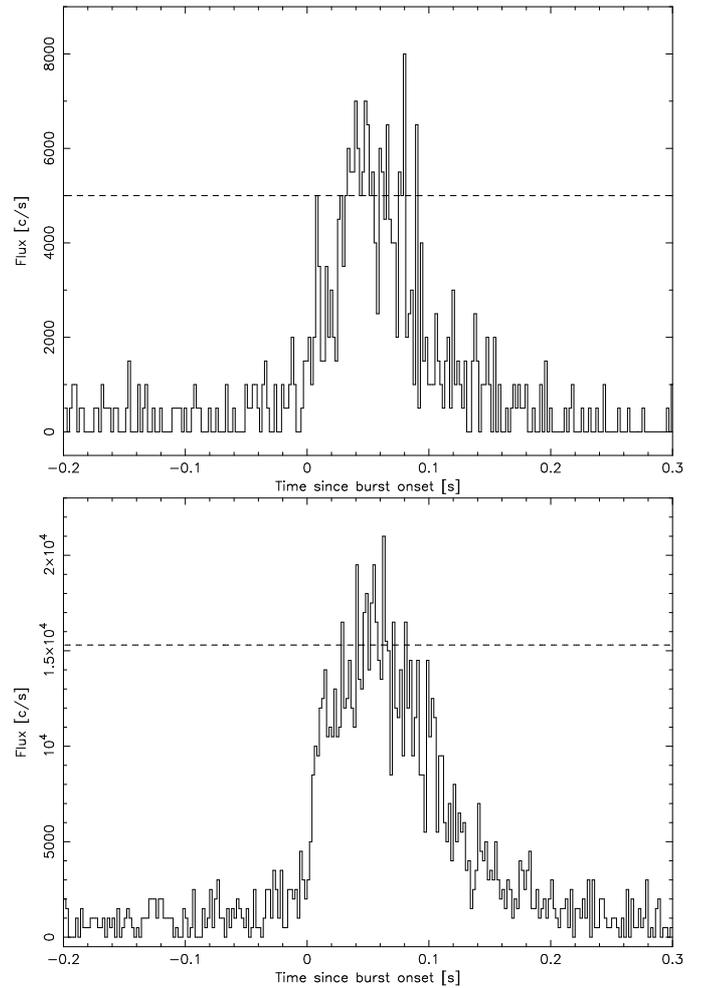

\includegraphics[height=0.98\columnwidth,angle=270]{precursors_fA1a.ps}
\includegraphics[height=0.98\columnwidth,angle=270]{precursors_fA1b.ps}
\caption{Full bandpass light curves of two bursts detected from 4U
  1820-30 on MJD 54958.740 (top; one active PCU) and MJD 54981.187
  (bottom; three active PCUs). The horizontal dashed lines indicate the
  Eddington limit in c~s$^{-1}$ as measured through the peak flux
  during the main burst phase later on. The time resolution of both
  light curves is 2 ms.
\label{fig:1820}}
\end{figure}

The shortest precursor after those presented in this paper was found
in a burst from 4U 1820-30, which was published in \cite{zan12} as
burst no. 4 (MJD 54958.740; ObsID 94090-02-02-01). The light curve is
shown in Fig.~\ref{fig:1820}. The rise time is $30\pm2$ ms. Another
higher quality burst is shown in the same figure. This also has a
short rise time and a somewhat longer duration. It is burst no. 7 (MJD
54981.187; ObsID 94090-01-05-00).

\section{PCA dead time correction for high time resolution}
\label{sec:deadtime}

\begin{figure}[t]
\includegraphics[height=0.98\columnwidth,angle=270]{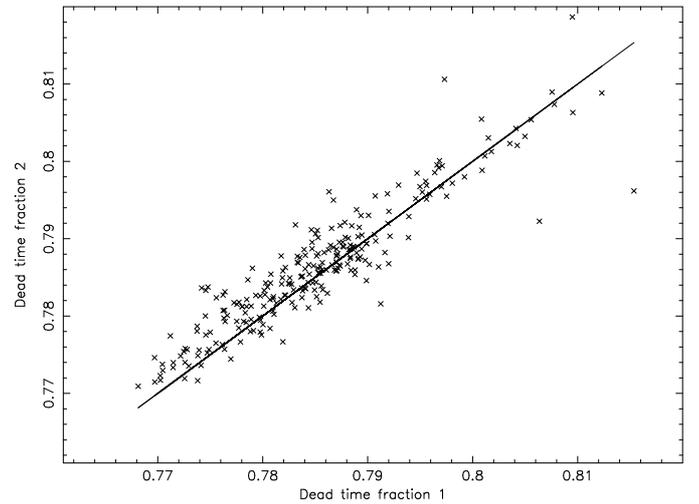}
\caption{Live time fraction '2' calculated from
  $0.98(1-1.25\times10^{-5}C)$ against live time fraction '1' calculated
  according to the formal recipe $1-(10^{-5}*C1+1.5\times10^{-4}*C2)$. The line
  indicates where both are equal.
\label{fig:deadtime}}
\end{figure}

The formal recipe for dead time correction\footnote{see the RXTE Cook
  Book at URL {\tt
    http://heasarc.nasa.gov/docs/xte/recipes/cook\_book.html}} uses
0.125~s resolution standard-1 data and is
$1/(1-10^{-5}*C1-1.5\times10^{-4}*C2)$ with $C1$ the combined rate per
PCU of the Good Xenon Events and Propane events and the coincidence
('Remaining') Events. The symbol $C2$ represents the count rate of the
Very Large Events (i.e., those that trigger the upper energy
discriminator). Unfortunately, this is not useful for the 10$^3$ times
larger time resolution employed here. Therefore, we calculated an
alternative dead time correction from only Good Xenon Events, since
these dominate the dead time during the bursts, through
$1/0.98(1-1.25\times10^{-5}C)$ with $C$ the Xenon event rate per PCU
from the event mode data. This alternative recipe was calibrated
against the formal recipe for a time resolution of 0.125~s throughout
the burst, see Fig.~\ref{fig:deadtime}. For \bone\ at 122 $\mu$s
resolution, the dead time fraction rises to 35\% at the peak of the
precursor, compared to about 20\% at the peak in the main burst, and
for \btwo\ to 22\%. Without dead time correction, the observed
precursor peak count rate is $2.0\pm0.1$ times that of the main burst
phase, for both bursts. After dead time correction, this rises to
$3.1\pm0.3$ for \bone\ and to $2.6\pm0.3$ for \btwo.

\end{document}